\documentclass[letter,twocolumn]{IEEEtran}
\usepackage{stmaryrd}
\usepackage{amssymb}
\usepackage{mathrsfs}
\usepackage{amsfonts}  
\usepackage[dvips,final]{graphicx}
\usepackage[dvips]{color}
\usepackage{epsfig,amssymb,amsmath}
\newcommand{\beq}{\begin{equation}}
\newcommand{\eeq}{\end{equation}}
\newcommand{\beqn}{\begin{eqnarray}}
\newcommand{\eeqn}{\end{eqnarray}}

\begin{document}

\title{Reduced Complexity Angle-Doppler-Range Estimation for  MIMO Radar That Employs Compressive Sensing}
\author{\authorblockN{Yao Yu, Athina P. Petropulu and H. Vincent Poor$^+$}\\
\authorblockA{  Electrical \& Computer Engineering Department, Drexel University\\
$^+$ School of Engineering and Applied Science, Princeton
University}
\thanks{This work was supported by the Office of Naval Research under Grant
ONR-N-00014-09-1-0342.} }
\maketitle

\begin{abstract}

The authors recently proposed a MIMO radar system that is
implemented by a small  wireless network. By applying compressive
sensing (CS) at the receive nodes, the MIMO radar super-resolution
can be achieved with far fewer observations than conventional
approaches. This previous work considered  the estimation of
direction of arrival and Doppler. Since the targets are sparse in
the angle-velocity space, target information can be extracted by
solving an $\ell_1$ minimization problem. In this paper, the range
information is exploited by introducing step frequency to MIMO radar
with CS. The proposed approach is able to achieve  high range
resolution and also improve the ambiguous velocity. However,
joint angle-Doppler-range estimation requires  discretization of the
angle-Doppler-range space which causes a  sharp rise in the
computational burden of  the $\ell_1$ minimization problem. To
maintain an acceptable complexity, a
 technique is proposed to successively estimate  angle,
Doppler and range in a decoupled fashion. The proposed approach can  significantly reduce
the complexity without sacrificing performance.

\end{abstract}

\section{Introduction}
 Multiple-input multiple-output (MIMO)
radar systems have received considerable attention  in recent years.
Unlike a phased-array radar, a MIMO radar \cite{Fishler:04}
transmits multiple independent waveforms from its antennas. MIMO
radar with widely separated antennas \cite{Haimovich: 08} exhibits
 spatial
diversity, which improves  target resolution. In colocated
MIMO radar \cite{Stoica: 07m}, phase differences induced by
 transmit and receive
antennas can  be exploited to form a long virtual array and thus achieve superior spatial resolution as
compared to traditional radar systems. Compressive sensing (CS)
is a recent development \cite{Donoho:06}-\cite{Candes:08}  and has already been
applied successfully in diverse fields such as image processing
 and wireless communications.  CS theory states that a $K$-sparse signal $\mathbf{x}$ of
length $N$ can be recovered exactly with high probability from
$\mathcal{O}(K\log N)$ measurements via $\ell_1$-optimization.

The application of  CS to  radar systems was
 investigated in \cite{Baraniuk:07}-\cite{Herman:08}, and  to  MIMO radar in
\cite{Petropulu:08}-\cite{Yu:09_tsp}. In \cite{Chen:081}, a uniform
linear array was considered as a transmit and receive antenna
configuration.  In \cite{Petropulu:08} and \cite{Yu:09_tsp}, the authors
proposed a MIMO radar implemented by a small scale network. According to \cite{Yu:09_tsp},  spatially distributed
network nodes,  each equipped with a single antenna, serve as transmit and receive antenna elements.
The transmit nodes transmit periodic pulses. The receive nodes forward
their compressive measurements to a fusion center. Exploiting the
spareness of targets  in the
 angle-Doppler space, an $\ell_1$-optimization problem is formulated and
solved at the fusion center in order to extract  target angle and
Doppler information. This approach achieves the superior resolution
of MIMO radar with far fewer samples than are required by
conventional approaches. This implies low power consumption for the
receive nodes.

The contribution of this paper is a low complexity CS approach for
obtaining range as well as direction of arrival (DOA) and Doppler
information about the target. Range estimation cannot not be
obtained  with the method of
 \cite{Yu:09_tsp}, as the target range  causes an identical
phase shift  to the signals received at all nodes and during all
pulses.
  One way to obtain range information
is to measure the travel time of the emitted radar signal
\cite{Chen:081}.  However,  the  range resolution of such an
approach might be limited. In this paper, we modify the transmitted
waveforms of the scheme of  \cite{Yu:09_tsp}, so that range
information is reflected in different phase shifts over different
pulses.
 In particular, the transmit nodes transmit pulses whose frequency increases by some step from one pulse to the next.
In this way, we introduce the step frequency radar approach
\cite{Gill} to MIMO radar, an approach that results in high range
resolution. Range resolution tends to increase as the transmitted
signal bandwidth increases. However, wideband signals correspond to
very short pulses, which experience low signal-to-noise ratios at
the receiver. A step frequency radar transmits a sequence of longer
pulses that are narrowband and  together cover a wide frequency
range. Thus, a step frequency radar  transmits an effectively
wideband signal that does not suffer from low SNR problems. An
extension of \cite{Yu:09_tsp} to joint angle-Doppler-range
estimation would be  straightforward; however, it would involve
prohibitively high complexity. In this paper, we propose an approach
to obtain angle-Doppler-range information in a decoupled fashion,
which results in significant complexity reduction. In
\cite{Yu:RWS2010}, a decoupled angle-Doppler estimation approach was
proposed for  the case of slowly moving targets. In that case,   we
can assume that the Doppler shift varies between pulses but within
each pulse remains approximately constant. Based on compressively
collected observations during one pulse, one can  obtain  initial
estimates of  the  azimuth angles by discretizing the angle space
only. Then, Doppler information is extracted by combining the data
of multiple pulses. The basis matrix requires a discretization of
the Doppler space only for the initial angle estimates. In this
paper, we apply a similar idea to decouple the estimation of angle,
Doppler and range. We propose to transmit a pulse train with
constant carrier frequency, followed by a pulse train with carrier
frequency that varies between pulses. Based on the received data
during the first pulse train one can decouple angle and Doppler
estimation along the lines of \cite{Yu:RWS2010}. Based on these
initial angle-Doppler estimates, the range information can be
extracted from the data corresponding to pulses that have varying
frequency. The proposed method significantly reduces the complexity
as compared to the joint angle-Doppler-range estimation using CS
without suffering performance degradation.

\section{Signal model for the constant carrier frequency}\label{sig_mod}
Let us consider the same setting as in \cite{Yu:09_tsp}. Assume $K$
point targets and colocated antennas randomly distributed in a small
area. The nodes transmit periodic pulses.
 The $k$-th target is at azimuth angle $\theta_k$ and moves
with constant radial speed $v_k$. Let $(r^t_{i},
\alpha^t_{i})$/$(r^r_{i}, \alpha^r_{i})$   denote the location of
the $i$-th transmit/receive node   in polar coordinates. The number
of transmit nodes  and receive nodes is denoted by $M_t$ and $N_r$,
respectively. Let $d_k(t)$ denote the range of the $k$-th target at
time $t$.
 Under the far-field assumption, i.e.,
 $d_{k}(t) \gg r^{t/r}_{i}$, the distance between the $i$th transmit/receive
node  and the $k$-th target
 $d^t_{ik}$/$d^r_{ik}$ can be approximated as
\begin{eqnarray}
d^{t/r}_{ik}(t) \approx d_k(t)- \eta_i^{t/r}(\theta_k)
=d_k(0)-\eta_{i}^{t/r}(\theta_k)-v_kt
\end{eqnarray}
where
$\eta_{i}^{t/r}(\theta_k)=r^{t/r}_{i}\cos(\theta_k-\alpha^{t/r}_{i})$.

Assuming that there are $N_j$ jammers  located at
$(\tilde{d}_j,\tilde{\theta}_j)$ and there is no clutter, the
compressive samples collected by the $l$-th antenna during the
$m$-th pulse  are given by
\begin{align}\label{rec_sig}
{\bf r}_{lm}&=
\sum_{k=1}^{K}(\tilde{\mathbf{\Phi}}_l\gamma_{k}e^{j{2\pi}p_{lmk}}{\bf
D}(f_{k}){\bf X}{\bf
v}(\theta_k))\nonumber\\
&+\underbrace{\sum_{j=1}^{N_j}\tilde{\mathbf{\Phi}}_le^{-j\frac{2\pi(\tilde{d}_j-\eta^r_{l}(\tilde{\theta}_j))f}
{c}}\tilde{\beta}_j
            \tilde{\bf x}_{jm}+\tilde{\mathbf{\Phi}}_l{\bf
e}_{lm}}_{{\bf y}_{lm}}
\end{align}
where
\begin{enumerate}
\item $c$ and $f$ denote the light speed and carrier frequency,
respectively; $T$ is the radar pulse repetition interval;
\item $\gamma_{k}=\beta_k e^{-j\frac{4\pi d_k(0)f}{c}}$;
$\beta_{k}$  denotes the reflection coefficient of the $k$-th
target;
\item $p_{lmk}=\eta_{l}^{r}(\theta_k)f/c+f_k(m-1)T$; $f_k=\frac{2v_kf}{c}$ is the doppler shift induced by the $k$-th target;
\item $lT_s,
l=0,\ldots,L-1$ represents the time within the pulse (fast time) and
thus the pulse duration is $LT_s$;
\item The $i$-th column of ${\bf X}$ contains the transmit
waveforms of the length $L$ from the $i$-th transmit node, where  ${\bf
X}^H{\bf X}={\bf I}_{M_t}$; 
\item $\tilde{\mathbf{\Phi}}_l=\mathbf{\Phi}_l{\bf X}^H \ (M\times L)$
 is the measurement matrix for the $l$-th receive node \cite{Yu:09_tsp};
 $\mathbf{\Phi}_l$ is an $M\times M_t \ (M\leq M_t)$ zero-mean
Gaussian random matrix;
\item ${\bf
v}(\theta_k)=[e^{j\frac{2\pi
f}{c}\eta^t_{1}(\theta_k)},...,e^{j\frac{2\pi
f}{c}\eta^t_{M_t}(\theta_k)}]^T$
 and ${\bf D}(f_k)={\rm
diag}\{[e^{j{2\pi}f_k0T_s},\ldots,e^{j{2\pi}f_k(L-1)T_s}]\}$;
\item $\tilde{{\bf x}}_{jm}$ denotes the waveform emitted by the $j$-th jammer during the $m$-th pulse; ${\bf e}_{lm}$ is the  thermal noise at the
$l$-th receive node corresponding to the $m$-th pulse; and the
corresponding powers are   $\tilde{\beta}^2_j$ and $\sigma^2$.
\end{enumerate}
It can be easily seen that the phase term associated with the range
$e^{-j\frac{4\pi d_k(0)}{\lambda}}$ is independent of receivers and
pulses and thus it can be absorbed by the reflection coefficient.

\section{Introduction of step frequency to MIMO radar using
CS}\label{mod_sf}

Let us consider a MIMO radar system in which each transmit node
transmits pulses, each of pulse repetition interval $T$,  so that
the carrier frequency of the $m$-th pulse equals
\begin{align}
f_m=f(1+\Delta f_m)
\end{align}
where $\Delta f_m$ is the frequency step, with $0<\Delta f_m<1, m=1,\ldots, N_p$.

The  baseband samples collected by the $l$-th antenna during the
$m$-th pulse are given by
\begin{align}\label{rec_sig_sf}
\tilde{{\bf r}}_{lm}&= \sum_{k=1}^{K}\gamma_ke^{j2\pi
\tilde{p}_{lmk}}{\bf \Phi}_l{\bf X}^H{\bf D}(f_{mk}){\bf X}{\bf
v}_m(\theta_k)+{\bf y}_{lm}
\end{align}
where
\begin{align}
&\ f_{mk}=\frac{2v_k f_m}{c}, {\bf v}_m(\theta_k)=[e^{j\frac{2\pi
f_m}{c}\eta^t_{1}(\theta_k)},...,e^{j\frac{2\pi
f_m}{c}\eta^t_{M_t}(\theta_k)}]^T\nonumber\\
&\tilde{p}_{lmk}=\frac{-2d_k(0)
f_m}{c}+\frac{\eta_{l}^{r}(\theta_k)f_m}{c}+{f_{mk}(m-1)T}.
\end{align}
In (\ref{rec_sig_sf}), the phase term associated with $d_k(0)$
varies with the pulse index.

 Let us
discretize the angle-velocity-range space on a fine grid: $
\mathbf{a}=[(a_1,b_1,c_1),\ldots,(a_N,b_N,c_N)]. $ Then
(\ref{rec_sig_sf}) can be rewritten as
\begin{align}\label{received signal}
\tilde{{\bf r}}_{lm}&= \sum_{n=1}^{N}s_ne^{j2\pi {q}_{lmn}}{\bf
\Phi}_l{\bf  X}^H{\bf D}(\frac{2b_nf_m}{c}){\bf X}{\bf
v}_m(a_n)+{\bf y}_{lm}
\end{align}
where
$
 s_n = \left\{
\begin{array}{rl}
\gamma_{k},  &  \text{if the $k$-th target is  at}\ (a_n,b_n,c_n) \\
0,  & \text{otherwise}
\end{array} \right.$ and
\begin{align}
q_{lmn}=\frac{-2c_n
f_m}{c}+\frac{\eta_{l}^{r}(a_n)f_m}{c}+\frac{2b_nf_m(m-1)T}{c}.
\end{align}
 In a compact matrix form we have $ \tilde{{\mathbf
r}}_{lm}=\mathbf{\Psi}_{lm}{\mathbf{s}}+{\bf y}_{lm} $, where
\begin{align}\label{sensing_matrix}
\mathbf{\Psi}_{lm}={\bf \Phi}_l{\bf  X}^H[&e^{j2\pi q_{lm1}}{\bf
D}(2b_1f_m/c){{\bf X}}{\bf v}_m(a_1),\ldots,\nonumber\\ &e^{j2\pi
q_{lmN}}{\bf D}(2b_Nf_m/c){{\bf X}}{\bf v}_m(a_N)].
\end{align}

 If
there are only a small number of targets as compared to $N$,  the
positions of targets are sparse in the angle-velocity-range space,
i.e., $\mathbf{s}$ is a sparse vector. A fusion center can combine
the compressively sampled signals due to $N_p$ pulses obtained at
$N_r$ receive nodes as
\begin{eqnarray}\label{cs}
\tilde{{\bf r}}&=&[\tilde{{\bf r}}^T_{11},\ldots,\tilde{{\bf
r}}^T_{1N_p},\ldots,\tilde{{\bf
r}}^T_{N_rN_p}]^T=\mathbf{\Theta}\mathbf{s}+{\bf Y}
\end{eqnarray}
where $\mathbf{\Theta}=[(\mathbf{\Psi}_{11})^T,\ldots,(
\mathbf{\Psi}_{1N_p})^T,\ldots,(\mathbf{\Psi}_{N_rN_p})^T]^T$ and
${\bf Y}=[({\bf y}_{11})^T,\ldots, ({\bf y}_{1N_p})^T,\ldots, ({\bf
y}_{N_rN_p})^T]^T$. The vector   $\mathbf{s}$ can be recovered by
applying the Dantzig selector \cite{Candes:07} to  (\ref{cs}). The
location of the non-zero elements of ${\bf s}$ provides information
on target angles, velocity and range.

\subsection{Unambiguous range and velocity}
In this section, we discuss the effects of step frequency on the
unambiguous range $R_u$ and unambiguous velocity $V_u$. In the case
of slowly moving targets, i.e., $f_{mk}T_sL<<1,\ k=1,\ldots, K,\
m=1,\ldots,N_p$, the Doppler shift change over the pulse duration
$T_sL$ is negligible as compared to the change between pulses.
 Consider
two  grid points $(a_i,b_i,c_i)$ and $(a_j,b_j,c_j)$ in the
angle-velocity-range space.  Given $a_i=a_j$, $b_i=b_j$ and $c_i\neq
c_j$, there is no range ambiguity if $e^{-j4\pi c_i f_m/c}\neq
e^{-j4\pi c_j f_m/c}, m=1, \ldots, N_p$. It holds that
\begin{itemize}
\item If $\Delta f_m=0$, then
$R_u=\frac{cT}{2}$;
\item If $\Delta f_m=(m-1)\Delta f$, then $R_u=\frac{c}{2f\Delta f}$ \cite{Gill};
\item If $\Delta f_m$ is randomly generated within a  predetermined range $[f_{min},f_{max}]$, then
$R_u\rightarrow \infty$ when $m$ is large.
\end{itemize}

Similarly, let $a_i=a_j$, $b_i\neq b_j$ and $c_i=c_j$.  If $e^{j4\pi
b_if_m(m-1)T/c}= e^{j4\pi b_jf_m(m-1)T/c}, m=1,\ldots,N_p$, then
velocity ambiguity will arise. Thus
\begin{itemize}
\item If $\Delta f_m=0$, then
$V_u=\frac{c}{2fT}$;
\item If $\Delta f_m=(m-1)\Delta f$, then
 $V_u $  equals the minimum common multiple of $\{\frac{c}{2f_mT}, m=1,\ldots
N_p\}$;
\item If $\Delta f_m$ is  randomly chosen from $[f_{min},f_{max}]$, then
 $V_u\rightarrow \infty$
when $m$ is large.
\end{itemize}

\subsection{Velocity resolution}
 Next we investigate the effects of step frequency on the velocity resolution in terms of  the column
correlation in the sensing matrix. To simplify the analysis, we
consider only one receive node. The sensing matrix for the $l$-th
receive antenna  is $
\mathbf{\Theta}_l=[\mathbf{\Psi}^T_{l1},\mathbf{\Psi}^T_{l2}\ldots,
         \mathbf{\Psi}_{lN_p}]^T
$, where $\mathbf{\Psi}_{lm}$ is defined in (\ref{sensing_matrix}).

On letting ${\bf g}_k$ denote the $i$-th column of
$\mathbf{\Theta}_l$, the correlation of columns ${\bf g}_k$ and
${\bf g}_{k'}$ equals
\begin{align}\label{CAGV}
&p_{kk'}=|<{\bf g}_k,{\bf g}_{k'}>|\nonumber\\
&=\left\{
\begin{array}{rl}
|\sum_{m=1}^{N_p}{\bf v}_m^H(a_k){\bf B}^{kk}_{lm}{\bf
v}_m(a_{k}) |& k=k'\\
|\sum_{m=1}^{N_p}e^{j2\pi(q_{lmk'}-q_{lmk})} {\bf v}_m^H(a_k){\bf
B}^{kk'}_{lm}{\bf v}_m^H(a_{k'}) |& k\neq k'
\end{array} \right.\
\end{align}
where  ${\bf B}^{kk'}_{lm}={\bf
    X}^H{\bf D}^H(\frac{2b_kf_m}{c})\tilde{\mathbf{\Phi}}_l^H\tilde{\mathbf{\Phi}}_l{\bf D}(\frac{2b_{k'}f_m}{c}){\bf
    X}$.

For simplicity, we make the following assumptions
\begin{itemize}
\item To highlight the velocity resolution, let $a_k=a_{k'}$ and
$c_k=c_{k'}$;
\item We  consider the correlation of columns corresponding to  the adjacent grid
points in the velocity dimension, i.e., $b_{k'}-b_k=\Delta b$. This
is the maximum correlation of columns in the velocity domain, and
thus dominates the velocity resolution. In this case, for slowly
moving targets, ${\bf B}^{kk'}_{lm}$ is approximately independent of
$\frac{2b_{k}f_m}{c}$;
\item Assume $\Delta f_m\ll 1$ and $r^t_i, i=1,\ldots,M_t$ are
sufficiently small. Then ${\bf v}_{m}(a_k)$ is approximately
identical across pulses.
\end{itemize}
Let  $\alpha=\frac{4\pi\Delta bTf}{c}$. Then (\ref{CAGV}) can be
approximated as
\begin{align}
&p_{kk'}\approx\nonumber\\
&\left\{
\begin{array}{rl}
{N_p}|{\bf v}_1^H(a_k) {\bf B}^{kk}_{l1} {\bf
v}_1(a_{k})|& k=k'\\
\underbrace{|\sum_{m=1}^{N_p}e^{j\alpha (1+\Delta
f_m)(m-1)}|}_{h(\mathbf{\Delta f})}|{\bf v}_1^H(a_k) {\bf
B}^{kk}_{l1} {\bf v}_1(a_{k})|& k\neq k'
\end{array} \right.\
\end{align}
where  $\mathbf{\Delta} f=[\Delta f_1,\ldots,\Delta f_{N_p}]$.

A set of sufficient conditions that guarantee a reduction in  the
correlation of columns corresponding to  adjacent grid points in the
velocity dimension, are the following (see proof in Appendix I):
\begin{itemize}
\item $\frac{\Delta f_{m}}{\Delta
f_{N_p+1-m}}\geq\frac{N_p-m}{m-1}$, $m>\lfloor N_p/2\rfloor$

(note that $\Delta f_m=m\Delta f$ satisfies this condition);
\item
$\sin(\alpha n)>0$ for $n=1,\ldots,N_p-1$.
\end{itemize}
For  example, for  $N_p=5$ the sufficient conditions require that
$\frac{\Delta f_{4}}{\Delta f_{2}}\geq\frac{1}{3}$ and
$\alpha<\frac{\pi}{4}$. It can be easily seen  that a larger
 $\Delta f_{m}-\Delta f_{N_p+1-m}, m>\lfloor N_p/2\rfloor$   can reduce
 the correlation of columns corresponding to adjacent grid
points on the velocity axis. On the other hand, as
 $\Delta f_m, m=1,\ldots,n_p$  increase the bandwidth consumption increases.
 Therefore, there is the tradeoff between the  velocity resolution and bandwidth.
 When $\Delta f_m\ll 1$, the gain in the velocity resolution due to the
 introduction of step frequency is negligible.

\section{Decoupled estimation of angle,
velocity and range  using CS}
\begin{figure}[htbp]
  \centering
    \includegraphics[height=3.2in,width=2.8in,clip=true]{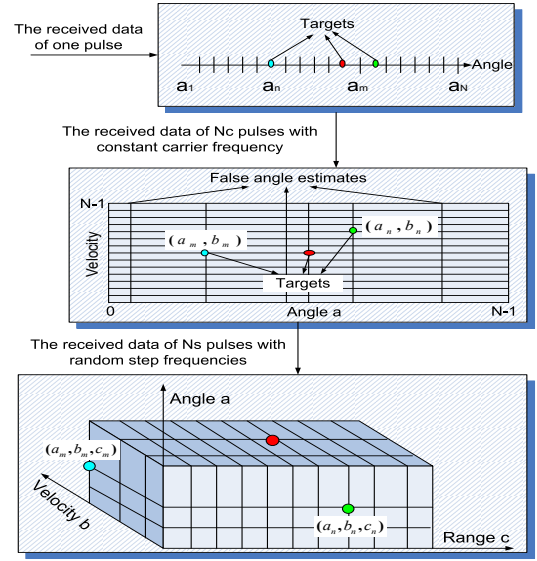}\caption{Illustration of the decoupled angle-velocity-range estimation
    approach
 }\label{decouple_scheme}
\end{figure}
Solving  an $\ell_1$ minimization problem requires polynomial time
in the dimension of $\mathbf{s}$. Let us discretize the angle,
Doppler and range space  into $[a_1,\ldots,a_{N_a}]$,
$[b_1,\ldots,b_{N_b}]$ and $[c_1,\ldots,c_{N_c}]$, respectively. The
joint estimation of angle, velocity and range requires complexity of
 $\mathcal{O}((N_aN_bN_c)^3)$. For large $N_a$, $N_b$ and $N_c$, the  complexity cost of the CS approach would be prohibitive.
In this section, we propose a decoupled angle-velocity-range
estimation approach to lower the complexity burden.  Let us consider
the case of slowly moving targets, i.e., $f_kT_sL\ll 1$. In this
case, the Doppler shift within a pulse can be ignored. We propose to
first transmit $N_c$ pulses with constant carrier frequency, and
then transmit $N_s$ pluses with random step frequency. Based on
these pulses, the estimation proceeds in the following steps.

{\textbf{Step 1}}:  For the data collected during one pulse, the
phase terms associated with range and Doppler are constant over all
receive nodes. Therefore, we can  estimate the azimuth angle by
discretization of the angle space only, as illustrated in the top
graph of Fig. \ref{decouple_scheme} and described in
\cite{Yu:RWS2010} in detail. The same process can be repeated on a
number of different received pulses. Let $\Gamma_i$ denote the set
of angle estimates obtained based on the $i$-th pulse. The union of
the angle-estimate sets $\Gamma_1,\ldots,\Gamma_{N_p}$ are the angle
estimates that are provided to the next step.

\textbf{Step 2}: As described in Section \ref{mod_sf}, the range
information can be excluded from the basis matrix for the data
 with  constant carrier frequency.
Therefore, we can extract the Doppler information by applying CS to
the data collected during the first $N_c$ pulses. The corresponding
 basis matrix can be formed based on the above obtained initial angle estimates
and a discretization of the Doppler space as shown in the  middle
graph of Fig. \ref{decouple_scheme}.

\textbf{Step 3}: By processing the received signal of $N_s$ pulses
with the random stepped frequencies, we can  extract  range
information as described in Section \ref{mod_sf} with the important
difference that  only the range axis  needs to be discretized and
used along with
 the angle-velocity estimates produced in Step 2 (see the bottom graph of Fig. \ref{decouple_scheme}).

The complexity of Step 1, Step 2 and Step 3 are respectively
$\mathcal{O}(A(N_a)^3)$, $\mathcal{O}((BN_b)^3)$ and
$\mathcal{O}((CN_c)^3)$, where $A$, $B$ and $C$ are scalars much
smaller than $N_a$, $N_b$ and $N_c$. Therefore, the total complexity
of the decoupled scheme is
$\mathcal{O}(A(N_a)^3+(BN_b)^3+(CN_c)^3)$. For large $N_a$, $N_b$
and $N_c$, it holds that $\mathcal{O}(A(N_a)^3+(BN_b)^3+(CN_c)^3)\ll
\mathcal{O}((N_aN_bN_c)^3)$ which implies significant savings.

\section{Simulations}
We consider a MIMO radar system with the  transmit/receive nodes
uniformly distributed on a disk of radius $10$m. The center carrier
frequency is $f=5 GHz$ and the pulse repetition interval is
$T=\frac{1}{4000} s$.  Each transmit node uses orthogonal QPSK
waveforms. The received signal is
 corrupted by  zero mean Gaussian noise.  The SNR is set to $0$ dB. The SNR here is defined as the ratio of  power of
transmit waveform  to that of   thermal noise at a receive node. A
jammer is located at angle $7 \textordmasculine$ and transmits an
unknown Gaussian random waveform  with amplitude 60.
  The target reflection coefficients are
 all one.

 Fig. \ref{ambiguity} compares the performance of velocity estimation with constant carrier frequency and randomly stepped
 frequency. The target scenes shown in Fig.
\ref{ambiguity} are generated via $100$ independent and random runs.
The grey scale  represents the times a target detected by CS
occupies a particular grid point in the target scene. A lighter
color indicates a higher occurrence frequency of a target. To
highlight the velocity estimation, we consider an extreme case in
which three targets are moving  in the same direction of
$0\textordmasculine$ and have the same range at the initial time,
i.e. $R=1500m$. The radial velocities of the three targets are $170
m/s$, $175 m/s$ and $180 m/s$, respectively. The unambiguous
velocity for the constant carrier frequency $f=5GHz$ is $120m/s$.
The possible ambiguous estimates of three targets are $50 m/s$, $55
m/s$ and $60 m/s$, respectively. The number of transmit nodes and
receive nodes are $M_t=30$ and $N_r=5$, respectively. $M=30$
measurements are obtained at each receive node. The top graph shows
the true target scene. We consider the worst case for velocity
estimation in which the three targets are located at adjacent grid
points in the velocity domain. Therefore,  a single bright spot
appears in the target scene instead of three spots. The last three
graphs demonstrate the target scenes produced by the CS approach
with constant and step carrier frequency. It can be seen from Fig.
\ref{ambiguity} that the introduction of random step frequency can
eliminate the velocity ambiguity by using sufficiently many pulses
($N_p=10$). In sharp contrast, the use of constant carrier frequency
always yields the ambiguous estimates around $50m/s$ using
the same number of pulses. 

 Fig. \ref{decouple} shows the estimates of  target locations in the angle-velocity-range
 space using
the proposed decoupled method. Three targets are moving in the
directions of $\{-0.5\textordmasculine, 0\textordmasculine,
0.5\textordmasculine\}$. The radial velocity of the three targets
are $70 m/s$, $75 m/s$ and $80 m/s$, respectively. The corresponding
ranges are $1200 m/s$, $1250 m/s$ and $1200 m/s$.  We sample the
angle-velocity-range space by the increment
$(0.5\textordmasculine,5m/s,50m)$. The number of transmit nodes and
receive nodes are $M_t=30$ and $N_r=30$, respectively. $M=30$
measurements are obtained at each receive node. The transmitters
first send $N_c=5$ pulses with constant carrier frequency and then
$N_s=5$ pulses with randomly stepped frequency. The top and bottom
graphs  show respectively the true  target locations and the
estimates produced by the  decoupled angle-Doppler-range scheme in
100 random and independent runs. We can see that the information on
the three targets is exactly recovered in each of the $100$ runs. A
false target arises in only one out of 100 runs. The false target is
quite close to the first target, i.e.,
$(-0.5\textordmasculine,80m/s,1100m)$. The decoupled scheme requires
only $0.002\%$ of  the complexity of  joint estimation of angle,
velocity and range.

\section{Conclusion}
We have proposed a CS based MIMO radar approach for obtaining
angle-velocity-range estimates. First, we
have introduced a step frequency approach in each transmit node,
which not only achieves high range resolution but also improve the
unambiguous velocity.
 Further, a decoupled angle-velocity-range estimation scheme has been  proposed to alleviate the complexity burden of CS applied to the joint angle-velocity-range estimation.
The proposed scheme can dramatically reduce the computational cost
of CS and still achieve  good performance.

\centerline{\bf Acknowledgment}
The authors wish to thank Dr. Rabinder Madan for essential contributions to this work.
\bibliographystyle{IEEE}

\appendix
\section{Appendix A}
 For a
given pair $(k,k'), k\neq k'$, satisfying $a_k=a_{k'}$ and
$b_{k'}-b_k=\Delta b$, $h({\bf\Delta }f)$ is proportional to the
ratio of $p_{kk'}$ to $p_{kk}$, which reveals the effect of
 $\mathbf{\Delta} f$ on  the correlation of columns corresponding to  the adjacent grid
points in the velocity dimension. Instead of analyzing
$h(\mathbf{\Delta} f)$, we define another function for convenience
as
 follows:
\begin{align}
&C(\mathbf{\Delta} f)=h^2(\mathbf{\Delta}
f)=\sum_{m=1}^{N_p}\sum_{n=1}^{N_p}e^{j\alpha (
(m-n)+\Delta f_m(m-1)-\Delta f_n(n-1))}\nonumber\\
&=N_p+\sum_{m=1}^{N_p}\sum_{n=m+1}^{N_p}2\cos(\alpha ( (m-n)+\Delta
f_m(m-1)\nonumber\\
&-\Delta f_n(n-1)))
\end{align}

 $C(\mathbf{\Delta} f)$ can be expanded by the Taylor
series of the first order as
\begin{align}\label{taylaor}
&C(\mathbf{\Delta} f)\approx C(\mathbf{0})+\sum_{m=1}^{N_p}\Delta
f_{m}\frac{\partial C(\mathbf{\Delta} f)}{\partial \Delta
f_{m}}|_{{\bf \Delta }f={\bf 0}}\nonumber\\
&=C(\mathbf{0})-2\alpha \sum_{m=\lfloor
N_p/2\rfloor+1}^{N_p}\sum_{n=m-1}^{N_P+1-m}(\Delta
f_{m}(m-1)\nonumber\\&-\Delta f_{N_p+1-m}(N_p-m))\sin(\alpha n)
\end{align}

$\sum_{m=1}^{N_p}\Delta f_{m}\frac{\partial C(\mathbf{\Delta}
f)}{\partial \Delta f_{m}}|_{{\bf \Delta }f={\bf 0}}$ is required to
be negative for $C(\mathbf{\Delta} f)<C(\mathbf{0}), \mathbf{\Delta}
f\neq 0$. Therefore, sufficient conditions that guarantees
$C(\mathbf{\Delta} f)<C(\mathbf{0}), \mathbf{\Delta} f\neq 0$ are
\begin{itemize}
\item $\frac{\Delta f_{m}}{\Delta
f_{N_p+1-m}}\geq\frac{N_p-m}{m-1}$, $m>\lfloor N_p/2\rfloor$

($\Delta f_m=m\Delta f$ satisfies this condition);
\item
$\sin(\alpha n)>0$ for $n=1,\ldots,N_p-1$.
\end{itemize}

\begin{figure}[htbp]
  \centering
    \includegraphics[height=2.5in,width=3.2in,clip=true]{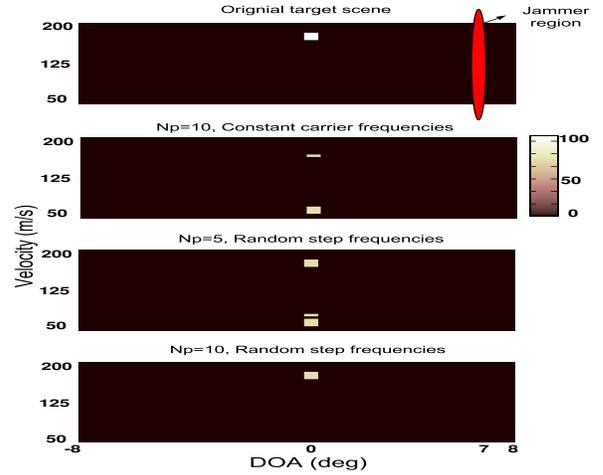}
    \caption{Velocity estimates using constant carrier frequency and randomly stepped frequency  for $M=M_t=30$ and $N_r=5$.
 }\label{ambiguity}
\end{figure}

\begin{figure}[htbp]
  \centering
    \includegraphics[height=2.8in,width=3.5in,clip=true]{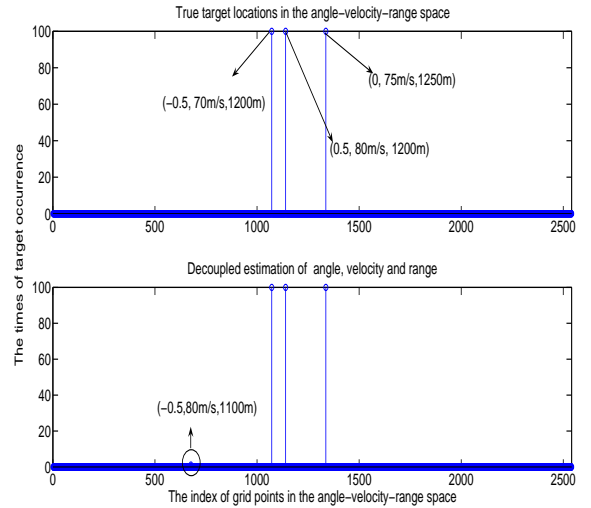}
    \caption{Target scenes produced by  decoupled angle-velocity-range estimation with randomly stepped frequency  for $M=M_t=N_r=30$ and $N_c=N_s=5$.
  }\label{decouple}
\end{figure}

\end{document}